\documentclass[fleqn,usenatbib]{mnras}
\usepackage{newtxtext,newtxmath}
\usepackage[T1]{fontenc}
\usepackage{ae,aecompl}
\usepackage{graphicx}
\usepackage{amsmath}
\usepackage{amssymb}
\usepackage{makecell}
\newcommand{\beq}{\begin{equation}}
\newcommand{\eeq}{\end{equation}}
\newcommand{\beqa}{\begin{equation}\begin{aligned}}
\newcommand{\eeqa}{\end{aligned}\end{equation}}

\newcommand{\Msun}{\, \rm M_\odot}
\newcommand{\hiMsun}{\,h^{-1}\rm M_\odot} 
\newcommand{\Mvir}{M_{\rm vir}}

\newcommand{\hikpc}{\,h^{-1}\rm kpc}
\newcommand{\hiMpc}{\,h^{-1}\rm Mpc}
\newcommand{\hiGpc}{\,h^{-1}\rm Gpc}

\newcommand{\Mpc}{\,{\rm Mpc}}
\newcommand{\Gpc}{\,{\rm Gpc}}

\newcommand{\Hloc}{H_0^{\rm loc}}
\newcommand{\dHloc}{\Delta H_0^{\rm loc}}
\newcommand{\HCMB}{H_0^{\rm CMB}}
\newcommand{\zmax}{z_{\rm max}}
\newcommand{\zmin}{z_{\rm min}}
\newcommand{\rmax}{r_{\rm max}}

\newcommand{\kmsMpc}{\,{\rm km\ s^{-1}Mpc^{-1}}}
\newcommand{\ds}{\displaystyle}
\newcommand{\nsnz}{n^\textrm{SN}(z)}
\newcommand{\nhaloz}{n^\textrm{halo}(z)}
\title[Sample variance in local measurements of $H_0$]{Sample variance in the local measurements of the Hubble constant}
\author[Wu and Huterer]{
Hao-Yi Wu$^{1}$\thanks{E-mail: hywu@caltech.edu} and
Dragan Huterer$^{2}$\thanks{E-mail: huterer@umich.edu} 
\\
$^{1}$California Institute of Technology, MC 367-17, Pasadena, CA 91125, USA \\
$^{2}$Department of Physics, University of Michigan, 450 Church Street, Ann Arbor, MI 48109, USA}
\date{Accepted 2017 July 28. Received 2017 July 28; in original form 2017 June 28}
\pubyear{2017}

\begin{document}
\label{firstpage}
\pagerange{\pageref{firstpage}--\pageref{lastpage}}
\maketitle

\begin{abstract}
The current $>3\sigma$ tension between the Hubble constant $H_0$ measured from
local distance indicators and from cosmic microwave background is one of the
most highly debated issues in cosmology, as it possibly
indicates new physics or unknown systematics.  In this work, we explore
whether this tension can be alleviated by the sample variance in the local
measurements, which use a small fraction of the Hubble volume.  We use a large-volume 
cosmological $N$-body simulation to model the local measurements and to
quantify the variance due to local density fluctuations and sample
selection. We explicitly take into account the inhomogeneous spatial
distribution of type Ia supernovae. Despite the faithful modelling of the
observations, our results confirm previous findings that sample variance in the
local Hubble constant ($\Hloc$) measurements is small; we find
$\sigma(\Hloc)=0.31\kmsMpc$, a nearly negligible fraction of the $\sim
6\kmsMpc$ necessary to explain the difference between the local and the global
$H_0$ measurements. While the $H_0$ tension could in principle be explained by
our local neighbourhood being a underdense region of radius $\sim 150$ Mpc, the
extreme required underdensity of such a void ($\delta\simeq -0.8$) makes it
very unlikely in a $\Lambda$CDM universe, and it also violates existing
observational constraints.  Therefore, sample variance in a $\Lambda$CDM
universe cannot appreciably alleviate the tension in $H_0$ measurements even
after taking into account the inhomogeneous selection of type Ia supernovae.

\end{abstract}


\section{Introduction}

The Hubble constant $H_0$ --- the current expansion rate of the Universe --- has
had a long history of increasingly precise measurements. It begins with Edwin
Hubble's groundbreaking but startlingly inaccurate measurement of
$500\kmsMpc$ \citep{Hubble29}, followed by subsequent decades with competing
but mutually discrepant claims that $H_0$ is either about $50\kmsMpc$
\citep[e.g.,][]{Sandage82} or about $100\kmsMpc$
\citep[e.g.,][]{deVaucouleurs79}. The situation was significantly clarified
with measurements that employed data from the Hubble Space Telescope, which
indicated $H_0=72\pm 8\kmsMpc$ (\citealt{Freedman01}; also see, e.g.,
\citealt{Freedman10}  for a review). Recent developments paved the way for truly
precision-level measurements of the Hubble constant --- one from the local
distance ladder and standard candles \citep[e.g,][]{Freedman12,Riess16}, and
the other from the global constraints on cosmological parameters using the cosmic
microwave background (CMB) measurements \citep[e.g.,][]{WMAP9,Planck15Cosmo}.

Interestingly, the latter two precise and physically very different types of
measurements currently appear to be in tension.  In particular, the $H_0$ derived
from the local distance ladder and type Ia Supernovae
\citep[SNe,][]{Riess09,Riess11,Riess16} has been significantly higher than the
$H_0$ derived from the acoustic peak scale of CMB
\citep[][]{Planck13Cosmo,Planck15Cosmo}.  For example, 
\citet[][R16 hereafter]{Riess16} presented $\Hloc = 73.24 \pm 1.74 \kmsMpc$, while the
Planck collaboration \citep[][P16 hereafter]{Planck1605.02985}
presented $\HCMB = 66.93 \pm 0.62 \kmsMpc$ (see their table 8).
The R16 value is $3.4\sigma$ higher, and the
relative difference, $(\Hloc - \HCMB) /\HCMB = 9\%$, is much larger than the
2.4\% error bar of R16 and the 0.9\% error bar of P16.

The $H_0$ discrepancy has caused much interest in the cosmology community, as
it could be a harbinger of new physics. Essentially, the data indicate that
the local measurement of the expansion rate, measured at distance scale
$\lesssim 400\Mpc$, is higher than the globally inferred value from the Hubble
volume $\sim(10\Gpc)^3$.  Various possibilities for this discrepancy have been
discussed, including the systematic errors in CMB \citep[e.g.,][]{Addison16}
or in the distance ladder \citep[e.g.,][]{Efstathiou14}, although the recent
independent analyses of the local measurements \citep{Cardona17,Feeney17,Follin17,Zhang17} confirm the R16
results.   Measurements using the time delay of gravitational lensing
\citep[e.g.,][]{Suyu13,Bonvin17} and the Tully-Fisher relation \citep[e.g.,][]{Sorce12}
appear to be consistent with the R16 local results, while measurements using
baryon acoustic oscillations \citep[e.g.,][]{Aubourg15} and tip of the red-giant 
branch distances \citep[e.g.,][]{Tammann13} are consistent with the P16
CMB results.

Local density fluctuations can make local measurements of the Hubble constant
deviate from the global value \citep[e.g.,][]{Turner92, Wang98, Shi98,
CoorayCaldwell06, HuiGreene06, Martinez-Vaquero09, Sinclair10, Courtois13,BenDayan14,Fleury17}.  In particular, if the Milky Way is located in an
underdense region in the cosmic web (the so-called ``Hubble bubble''), nearby
galaxies will tend to have positive peculiar velocities, which will bias the
$\Hloc$ measurement high. Several lines of evidence have supported the idea of
a local underdensity \citep[e.g.,][]{Zehavi98, Jha07, Keenan13,
WhitbournShanks14}.  However, it is unclear whether such a local underdensity
can fully account for the difference between $\Hloc$ and $\HCMB$.

Previous works have explored the statistical and systematic errors of $\Hloc$
using analytic models and $N$-body simulations.  \cite{Marra13} analytically
calculated the systematic error of $\Hloc$ due to local density perturbations.
They found that the local density perturbations can account for at most 2.4\%
of the systematic errors of $\Hloc$ measured at $0.01<z<0.1$, and that a
6$\kmsMpc$ deviation in $\Hloc$ is very rare.  Using cosmological $N$-body
simulations, \citet[][W14 hereafter]{Wojtak14} explored observers located in
voids and on dark matter haloes of different masses, quantifying the
deviation of $\Hloc$ as a function of distance scales.  They found that the
discrepancy between $\Hloc$ and $\HCMB$ cannot be accounted for using
density fluctuations, unless the Milky Way is located at the centre of one of the 10  
largest voids found in 6 $\hiGpc$ volume.  \cite{Odderskov14} also reached
the conclusion that the sample variance is too small to alleviate the tension,
after taking into account the depth and sky coverage of SN observations in light-cone simulations.

The purpose of this work is to explicitly incorporate in the analysis the
highly non-uniform spatial distribution of SNe used in the measurement of
$\Hloc$.  We quantify two sources of sample variance:\footnote{In this work we
use the two terms ``sample variance'' and ``cosmic variance''
interchangeably.}  local density contrast and inhomogeneous selection of SN
sample.  The aforementioned previous theoretical analyses of sample variance
in $\Hloc$ relied on various schemes of weighting and averaging radial
velocities of haloes isotropically selected from $N$-body simulations. However,
even with weighting schemes that are judiciously chosen to mimic the actual
observations, it is a priori far from obvious to us that the
inhomogeneity of the SN sample would not increase the sample variance
in the derived value of $\Hloc$.

Our approach is to explicitly mimic the highly inhomogeneous selection
function of SNe, using the R16 Supercal SN sample as a baseline. We incorporate
variations in how the R16 SN sample is orientated relative to the simulation's
coordinate system. We also study how haloes are assigned to physical SN
locations in order to use them to calculate $\Hloc $ in simulations. 
We note that R16 explicitly corrected for the peculiar velocity of
each SN using the density field from 2M++ \citep{Carrick15}; however, such
corrections are susceptible to large systematic uncertainties in the local
density field reconstruction.  To be conservative, we do not take into account
the fact that such corrections have been applied in the R16 analysis.
Therefore, our calculation corresponds to an upper limit of how much sample
variance contributes to the error budget; the sample variance will be smaller
if the peculiar velocity correction is sufficiently accurate.

This paper is organized as follows.  
Section~\ref{sec:method} describes our approach for quantifying the sample variance of $\Hloc$.
Section~\ref{sec:implementation} details how we map observed SNe to dark matter haloes.
Section \ref{sec:results} presents our main results, including 
the distribution of $\Hloc$ from various sample selections.
We explore the effect of density fluctuations in Section~\ref{sec:delta}.
We discuss our results in Section~\ref{sec:discussions} and conclude in Section~\ref{sec:conclusions}.
Throughout this work, we use the same cosmological parameters as the in the Dark Sky simulations (see Section~\ref{sec:darksky}).

\begin{figure}
\includegraphics[width=\columnwidth]{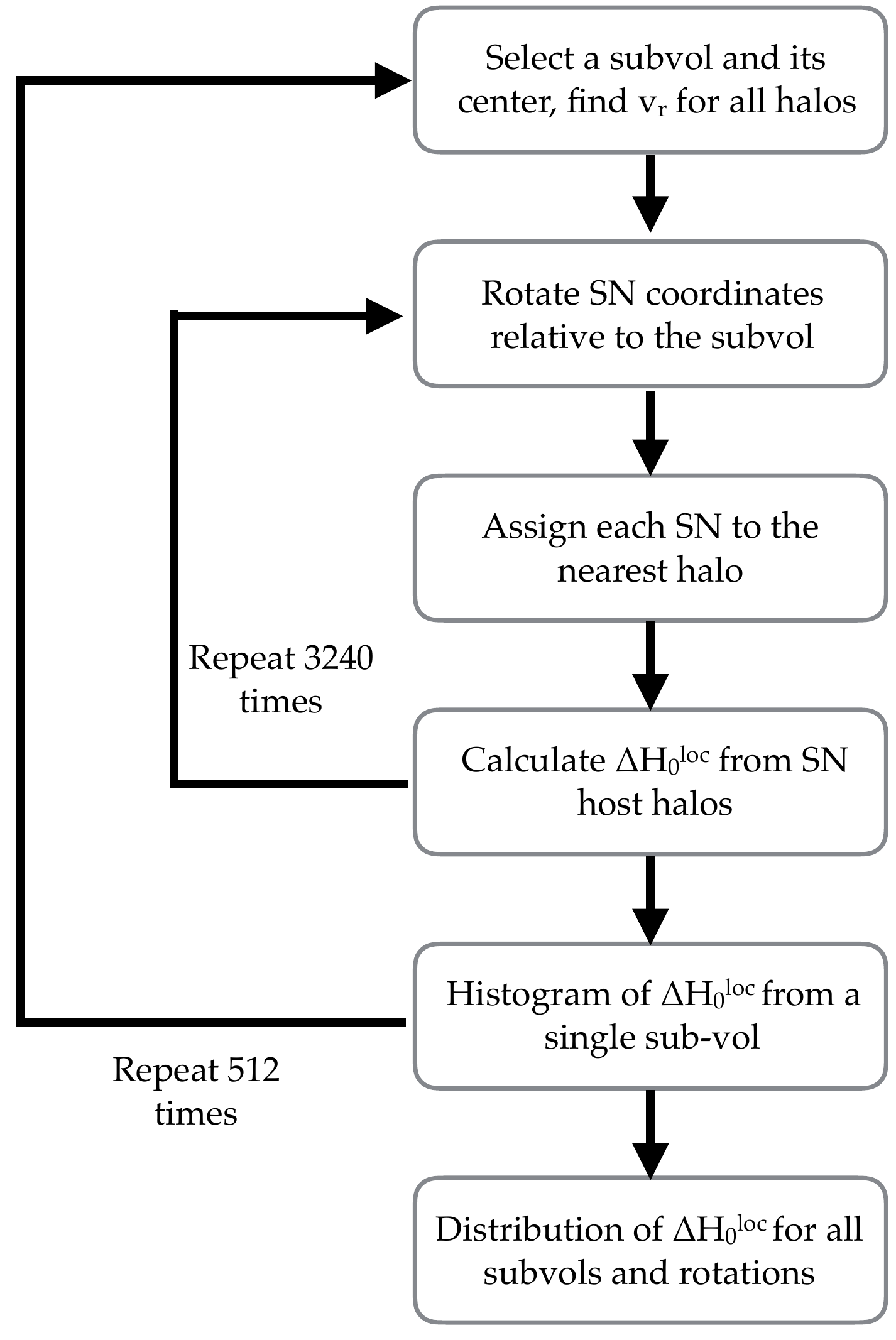}
\caption[]{
Flowchart depicting our simulation procedure. 
The inner loop runs over all mutual rotations between the SN and
the simulation coordinate systems. The outer loop runs over all
sub-volumes from the Dark Sky simulations, each sub-volume representing 
a realization of the local Universe.}
\label{fig:dH_flowchart}
\end{figure}

\section{Simulating the $\Hloc$ measurement}\label{sec:method}

In this section, we review how the $\Hloc$ measurement is done, introduce our
procedure for simulating this process, and derive the estimator for $\Hloc$.

\subsection{Overview of our approach}

We start by briefly reviewing how the Hubble constant is measured in 
R16.  The R16 $\Hloc$ measurement first employs four distance ``anchors'' with
direct geometric distance measurements\footnote{These anchors are: parallaxes
of Cepheids in the Milky Way, water masers in NGC 4258, and detached eclipsing
binaries in the Large Magellanic Cloud and in M31.}.  The second rung of the
distance ladder is a set of 19 galaxies (``calibrators'') with both Cepheids
and SNe;  the four anchors set the absolute distance of Cepheids,  and the 19
calibrators provide distance calibration for SNe.  The third and final rung of
the distance ladder is a set of $217$ SNe Ia at $0.023<z<0.15$.  The global
fit to the set of all standard candles --- anchors, calibrators, and SNe Ia
--- is used to determine the value of $\Hloc$.

When we use the distance--redshift relation of standard candles to derive
$\Hloc$, the redshift can be modified by the local density fluctuations, and
the derived $\Hloc$ is thus biased.  The redshift range adopted in R16, $0.023
< z < 0.15$, is selected so as to avoid the effect of peculiar velocities at
lower redshift and the effect of dark energy at higher redshift.

In this paper, we simulate the selection of the R16 SN distances
in order to account for the sample variance of $\Hloc$. 
Our approach is represented visually in Fig.~\ref{fig:dH_flowchart} and is
summarized as follows:  

\begin{enumerate}
\item Select a sub-volume in a cosmological simulation, within which each halo
in some mass range is a possible host of an SN.  

\item Orient the R16 SN coordinates with respect to the sub-volume.

\item Assign each of the SNe to the closest halo.

\item Calculate the deviation of the local Hubble constant, $\dHloc = \Hloc - H_0^{\rm true}$, from the radial velocities of these SN hosts.

\item Go to (ii), repeat the measurements for many different orientations, and 
obtain the histogram of $\dHloc$ of different orientations from a single sub-volume.

\item Go to (i), repeat the measurements for multiple, non-overlapping
  sub-volumes, and obtain the distribution of $\dHloc$ from all sub-volumes and
  all orientations.

\end{enumerate}

\subsection{From SN to $\dHloc$} \label{sec:dHloc}

Once the overall SN distance scale has been calibrated with nearby objects
(Cepheids and the four anchors), the redshift and apparent magnitude of each
SN provides a direct --- but noisy --- estimate of the Hubble constant. That
is, the magnitude fluctuation, $\delta m=(5/\ln 10)(\delta r/r)\propto v_r/r$,
is precisely an estimate of the deviation in the Hubble constant; 
here $v_r$ is the peculiar velocity along the line of sight, and $r$ is the comoving distance.
Below we adopt the formalism from R16 to drive the estimator of $\Hloc$.

In R16, the information from SNe is compressed into one number, $a_x$, which is
defined as the intercept of the magnitude--redshift relation in the $x$-band and
is related to $\Hloc$ via 
\beq \log_{10} \Hloc = \frac{M_x^0}{5} + a_x + 5 \ ,
\label{eq:ax}
\eeq
where $M_x^0$ is the absolute magnitude of SNe Ia derived from the distance anchors.

The value of $a_x$ is determined from the Hubble diagram (see fig.~8 in R16).
Let $m_{x,i}^0$ be the apparent magnitude of the $i^{th}$ SN in the $x$-band (fit by the SALT2 light-curve fitting algorithm) and $z_i$ its measured redshift; 
$a_x$ is then given by
\beq
a_x = \frac{1}{N} \sum_{i=1}^{N} \log_{10} \bigg(cz_i \big(1+O(z_i)\big)\bigg) - 0.2 m_{x,i}^0 \ .
\eeq
The measured redshift is perturbed by the peculiar velocity along the line of sight, which leads to 
a perturbation in $a_x$,
\beq
\Delta a_x \simeq \frac{1}{N} \sum_{i=1}^{N}  \frac{1}{\ln 10} \frac{c \Delta z_i}{cz_i}  
       \simeq \frac{1}{N} \sum_{i=1}^{N} \frac{1}{\ln 10} \frac{v_{r,i}}{r_i \Hloc} \, .
\label{eq:delta_ax}
\eeq
Here we adopt  $\Delta z = v_{r} / c$ and $r = cz/\Hloc$, which hold for $z \ll 1$.
From equations (\ref{eq:ax}) and (\ref{eq:delta_ax}), we obtain
\beq
\dHloc = (\Hloc \ln 10) \,\Delta a_x = \frac{1}{N} \sum_{i=1}^{N}\frac{v_{r,i}}{r_i}  \, .
\label{eq:dH0_averaged}
\eeq 
Equation (\ref{eq:dH0_averaged}) simply averages the
contribution from each SN, and it makes sense to instead inverse-variance weight their
contributions; this leads to the estimator we employ in this paper:
\beq
\dHloc =
\frac{\ds\sum_{i=1}^{N}\,\ds\frac{1}{\sigma_i^{2}}\,\ds\frac{v_{r,i}}{r_i}}
 {\ds\sum_{i=1}^{N}\,\ds\frac{1}{\sigma_i^{2}}}
 \qquad\mbox{(our estimator)}  .
\label{eq:dH0_weighted}
\eeq
Here $\sigma_i$ is the error bar of the magnitude of each SN, set to the square root of the diagonal of the noise covariance matrix of SNe \citep[see][]{Scolnic14}.

Since each SN gives a measurement of $v_r/r$, the minimum-variance estimator
simply averages those measurements with the inverse-variance weighting. Our
estimator in equation (\ref{eq:dH0_weighted}) is the appropriate one given our
goal to simulate the use of SNe in obtaining the Hubble constant.  We note
that there is an alternative estimator based on least squares (LSQ), $\sum
(w(r)r v_r)/\sum( w(r) r^2)$,  where $w(r)$ is the weight (see, e.g., W14 and
\citealt{Odderskov17})\footnote{We note that W14 used $v_r$  averaged in a
redshift bin and weighted each bin by $r^2$ (R. Wojtak, private
communication).  This procedure is equivalent to our equation
(\ref{eq:dH0_averaged}).}. Our estimator is equivalent to the LSQ estimator with each object
weighted by $r^{-2}$ (see Appendix \ref{app:weighting} for discussions on weighting). 
Therefore, our estimator gives more weight to objects at
small $r$.  For the distribution of SNe from R16 --- shown in Fig.~\ref{fig:nz}
--- our estimator returns a \textit{three times larger} standard deviation in
$\Hloc$ than the LSQ estimator; see Appendix \ref{app:estimators}. We will see
below that, despite giving a larger variance when applied to the R16 set of SNe, 
the sample variance in $\Hloc$ remains small when compared to the
difference between P16 and R16 measurements.

\section{Implementation}\label{sec:implementation}

In this section, we provide further details of the observation data and simulations used in our analysis. 

\subsection{Supercal SN data}\label{sec:supercal}
\begin{figure}
\includegraphics[width=\columnwidth]{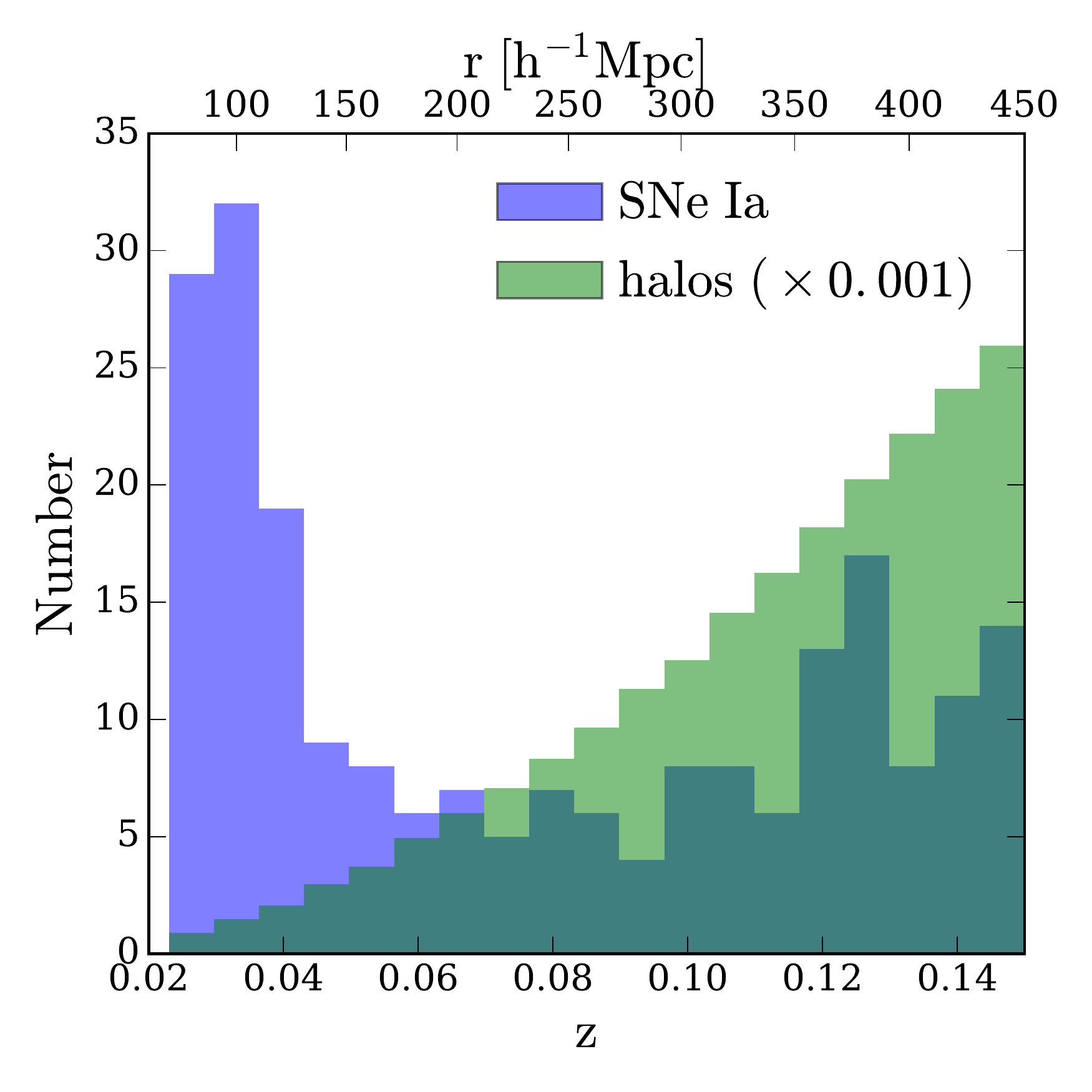}\\
\includegraphics[width=\columnwidth]{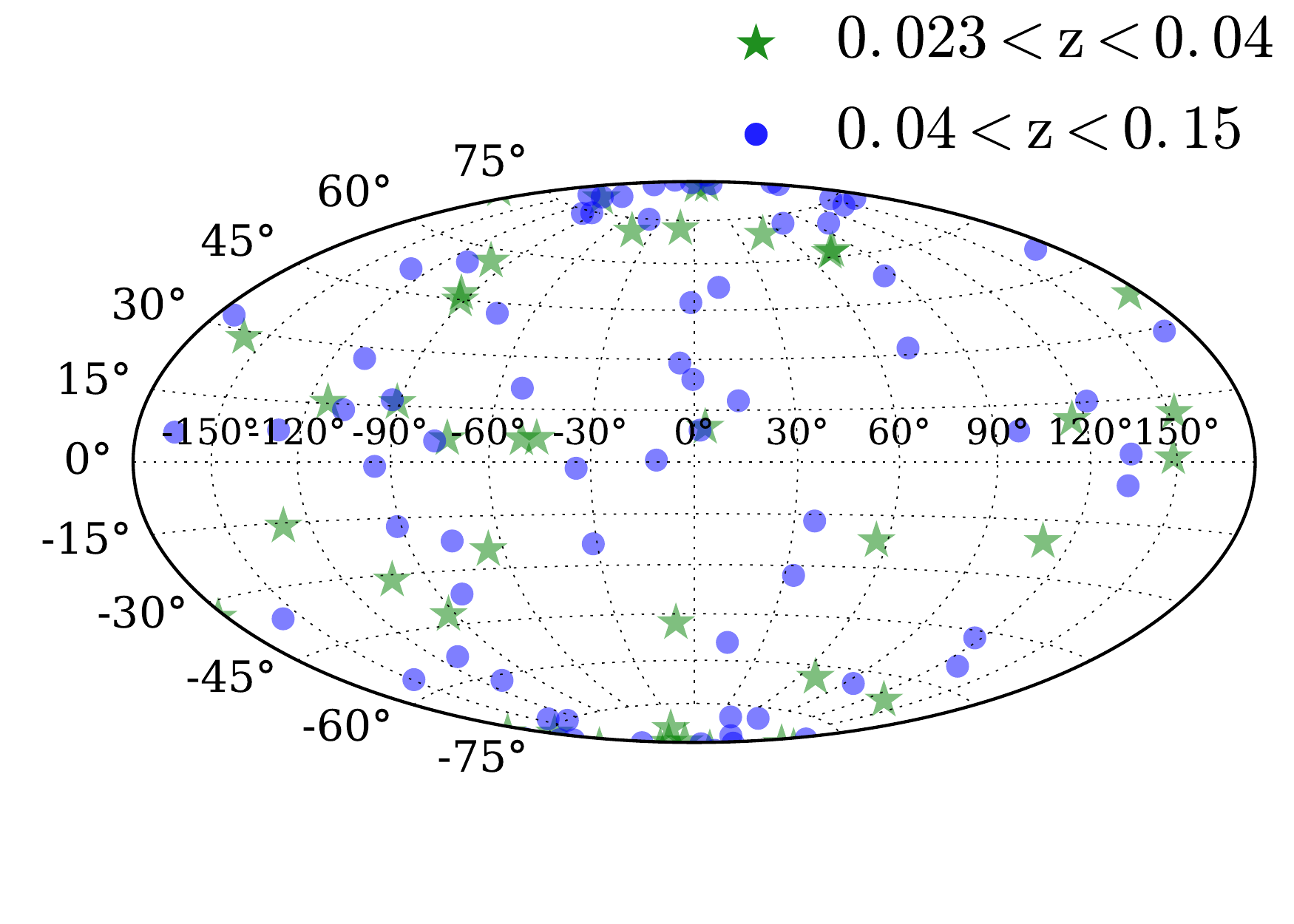}
\caption[]{ \textit{Top}: redshift distributions of the Supercal SN sample
  (blue histogram) and the dark matter haloes in Dark Sky simulations (green
  histogram), for $0.023 <z <0.15$, split into 20 bins.  \textit{Bottom}:
  angular distribution of the Supercal SN sample in the Galactic coordinates,
  shown with the Hammer (equal-area) projection.  The two symbols correspond to
  two ranges of redshift as shown.  Note that the angular distribution is
  highly sparse.}
\label{fig:nz}
\end{figure}

We use the recent compilation of SN data ``Supercal''
\citep{Scolnic14,Scolnic15}, which features a uniform photometric calibration
based on the consistent photometric system of Pan-STARRS1 across $3\pi$ of the
sky. This data set is based on the SALT2 light-curve model \citep{Guy10} and
includes a correction of distance bias due to intrinsic scatter of SN
brightness and measurement noise \citep{ScolnicKessler16}.  This data set is
also used in the recent $\Hloc$ analysis of R16 (see their section 4.3).   In
our analysis, we follow R16 and use 217 Supercal SNe Ia that lie in
$0.023<z<0.15$.  This choice of lower limit $z=0.023$ corresponds to $\simeq$
100 Mpc, above which the effects of peculiar velocities in SN data are small.
The upper panel of Fig.~\ref{fig:nz} shows the redshift distribution of these
217 SNe Ia; the radial distribution is skewed towards low redshift and is
different from a volume-limited sample, for which the number per bin would be
$\propto z^2$.

The lower panel of Fig.~\ref{fig:nz} shows the Hammer (equal-area) projection
of the positions of the Supercal SN sample in the Galactic coordinates.  While the
SNe are found over the whole sky, their overall distribution is
inhomogeneous. This becomes particularly obvious if we consider the full
three-dimensional distribution of SNe in space.

\subsection{Dark Sky simulations}\label{sec:darksky}

We use the public release of the Dark Sky
simulations\footnote{\href{http://darksky.slac.stanford.edu}{http://darksky.slac.stanford.edu}}
\citep{Skillman14}.  In particular, we use the largest volume {\tt ds14\_a}
with $1.07\times 10^{12}$ ($10240^3$) particles within a volume of $(8
\hiGpc)^3$.  The mass resolution is $3.9\times10^{10}\hiMsun$, and the Plummer
equivalent softening length is $36.8\hikpc$ .  The cosmological parameters correspond 
to a flat $\Lambda$CDM model and 
are consistent with \textit{Planck} and other probes  \cite[e.g.,][]{Planck15Cosmo}: 
$\Omega_M$ = 0.295;
$\Omega_b$ = 0.0468;
$\Omega_\Lambda$ = 0.705;
$h$ = 0.688;
$\sigma_8$ = 0.835.

The $N$-body simulation is performed using the adaptive tree code {\sc 2HOT}
\citep{Warren13}, and the dark matter haloes are identified using the halo
finder {\sc Rockstar} \citep{Behroozi13rs}.  The data is accessible online
using {\sc yt} \citep{Turk11}.  We use dark matter haloes with virial mass
$\Mvir > 10^{12.3}\Msun$ (35 particles).  We have explicitly verified the
completeness of haloes at this mass.

We divide this $(8\hiGpc)^3$ volume into 512 sub-volumes of $(1\hiGpc)^3$.  We
then choose the halo with virial mass $\Mvir \in [10^{12.3}, 10^{12.4}]\Msun$
that is closest to the centre of each sub-volume as our observer.  This choice
simulates 512 separate observers located on Milky Way mass haloes 
\citep[see e.g.,][for constraints on Milky Way mass]{Bland-Hawthorn16}, and each observer has
a separate sub-volume of the large-scale structure out to the distance of
interest ($\zmax = 0.15$). For the host haloes of SNe, we also use Milky Way
mass haloes with $\Mvir \in [10^{12.3}, 10^{12.4}]\Msun$, and we have
explicitly checked that this choice leads to the same results as using all
haloes above $10^{13}\Msun$.

Using $N$-body simulations, W14 found that if observers are located on dark
matter haloes, $\Hloc$ will be biased low for $\rmax \lesssim 100 \hiMpc$, and the
deviation is larger for higher halo mass.   In Appendix~\ref{app:weighting},
we show that our choice of halo mass leads to no bias on $\Hloc$ at all
scales. In addition, W14 showed that using the velocities with respect to the
CMB rest frame is the correct choice, while using the velocities with respect
to the observer's rest frame will lead to a unphysical bias.   We use halo
velocities with respect to the simulation's rest frame (which corresponds to
the CMB rest frame).

\subsection{Matching SNe to haloes}\label{sec:sn_to_halo}

The top panel of Fig.~\ref{fig:nz} shows the redshift distribution of the SNe
from R16 (blue histogram) and that of the dark matter haloes from the Dark Sky
simulations (green histogram).  While the number of haloes per redshift slice
scales with $z^2$, the number of SNe peaks at $z \simeq 0.03$. Moreover, there
are $\sim$200,000 haloes in each sub-volume in the aforementioned mass interval
$[10^{12.3}, 10^{12.4}]\Msun$ and redshift range $0.023<z<0.15$, but there are
only 217 SNe in this redshift range.  Below we show that the skewed and sparse
distribution of SNe will lead to additional sample variance.

The matching of SNe to haloes is done as follows. First, for each Dark Sky
sub-volume ($1\hiGpc$ on a side), we find the halo closest to the centre  (a
halo in our mass range can typically be found within $\sim15\hiMpc$ of any
given point). We place the observer on this halo, which is roughly the Milky Way mass 
given our choice for halo masses.

We then consider how to orientate the SN coordinate system relative to the
simulation sub-volume. While the distribution of SNe in redshift and angle is
given, the orientation of their coordinate system relative to that of the
simulation frame is arbitrary and, given the highly inhomogeneous distribution
of SN in the volume, will surely lead to additional variance. To account for
this previously unaccounted-for source of the sample variance, we explore many
possible orientations of the SN frame within a fixed simulation frame. To vary over
the orientations, we employ 3240 Euler angles; see Appendix~\ref{app:Euler}
for details. We explicitly verified that the corresponding angular step is sufficiently small to provide converged results.

Now that the three-dimensional coordinates of each SN and halo have been calculated 
for each rotation, we assign the closest halo to each SN.
The radial velocity of that SN is simply given by
\begin{equation}
v_{r,i}\equiv \textbf{v}_i\cdot
\frac{(\textbf{r}_i-\textbf{r}_{\rm obs})}{|\textbf{r}_i-\textbf{r}_{\rm obs}|} \ ,
\end{equation}
where the label $i$ refers to the $i^{\rm th}$ SN, 
and $\textbf{r}_{\rm obs}$ is the location of the observer.

\section{Results}\label{sec:results}

\begin{figure*}
\includegraphics[width=0.8\textwidth]{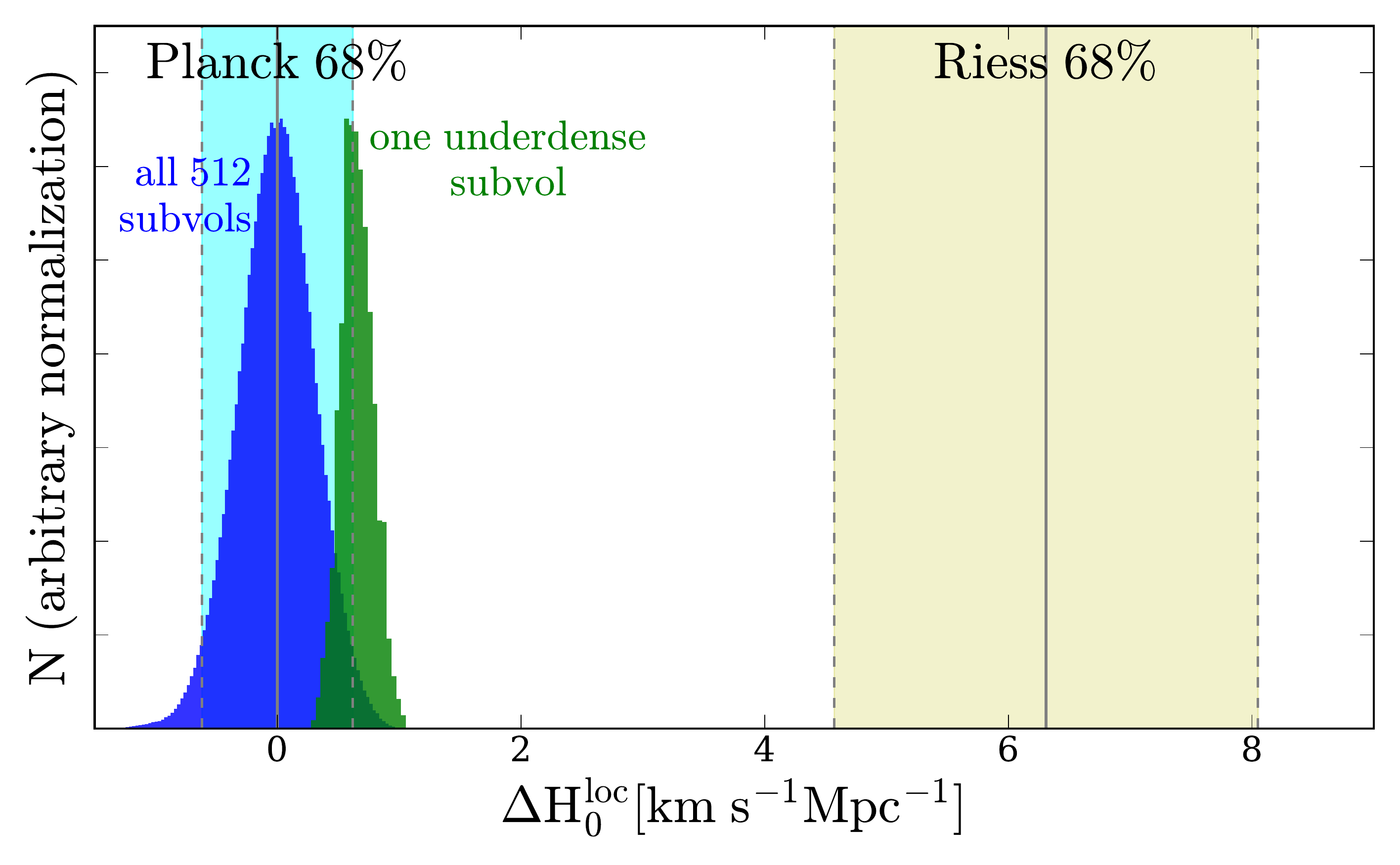}
\caption[]{Sample variance in $\dHloc$ from our simulations, compared to P16 and R16 error bars 
(assuming P16 is the true global value). The
blue histogram shows 3240 rotations of the SN coordinate system from 512 sub-volumes in the 
Dark Sky simulations, corresponding to $\sim$1.5 million
SN-to-halo coordinate system configurations. 
The green, slightly more jagged, histogram shows the results
of a particularly underdense sub-volume with a high $\dHloc$ at the 2-$\sigma$ level
relative to all sub-volumes. The two histograms are separately
normalized.  Note that the sample variance in $\Hloc$ is much smaller than
the difference between R16 and P16 measurements.  }
\label{fig:dH0_histograms}
\end{figure*}

\renewcommand{\theadalign}{cb}
\renewcommand{\theadfont}{\bfseries}
\renewcommand{\theadgape}{\Gape[4pt]}
\renewcommand{\cellgape}{\Gape[4pt]}
\renewcommand{\arraystretch}{1.6}

\begin{table*}
\begin{center}
\begin{tabular}{| *5{>{\centering\arraybackslash}m{1in}|}@{}m{0pt}@{}}
\hline
\multicolumn{5}{|c|}{Contributions to rms variance in $\dHloc$\quad ($\kmsMpc$)}\\
\hline\hline
\thead{Source of scatter:}         
& \makecell{Box-to-box,\\ all haloes, \\ no weighting} 
& + \makecell{weighted by\\ $\nsnz/\nhaloz$}
& + \makecell{3D matching, \\ 3240 rotations, \\ no weighting} 
& + \makecell{weighted by mag err \\(equation~\ref{eq:dH0_weighted})}   \\\hline
\thead{Cumulative\\[-0.05cm] $\sigma$ ($0.023 < z < 0.15$):}  
& $0.12$
& $0.38$
& $0.42$
& $\mathbf{0.31}$ \\\hline 
\end{tabular}
\end{center}
\caption{Contributions to the total sample variance by various assumptions
about the $\Hloc$ measurements.}
\label{tab:dH0}
\end{table*}

Our principal results are shown in Fig.~\ref{fig:dH0_histograms} and Table
\ref{tab:dH0}.

Fig.~\ref{fig:dH0_histograms} shows the distribution of $\dHloc$ from our
simulated measurements, compared with the best-fitting values and $1$-$\sigma$
measurement errors of R16 and P16.  Both histograms are separately normalized
for visual clarity.  The blue histogram includes 3240 rotations of the SN
coordinate system in each of the 512 sub-volumes, for a total of $\gtrsim$ 1.5
million configurations.  This histogram is approximately Gaussian with a
standard deviation of 0.31 $\kmsMpc$.  In units of this sample variance-only error, the
R16 measured value is about 19 standard deviations away from the P16 value. In
addition, the green histogram shows an example of a very underdense sub-volume,
with $\delta = -0.16$ (at the scale of 120 $\hiMpc$) and
$\dHloc=0.65\pm0.13\kmsMpc$ ($\sim$2$\sigma$ upward fluctuation). None of our
$\sim$1.5 million configurations comes even close to appreciably
helping reconcile the P16 and R16 measurements.

Table \ref{tab:dH0} shows the cumulative value of the standard deviation in
$\dHloc$ as various contributions to the sample variance are added to our
simulations. Using all haloes (without weighting) in the range $0.023<z<0.15$,
the mean of $\dHloc$ values among the 512 sub-volumes is less than 0.01
$\kmsMpc$,  and the standard deviation is 0.12 $\kmsMpc$.  Still calculating
$\dHloc$ from all haloes, but now weighting each halo's contribution with
$\nsnz/\nhaloz$, produces $\sigma(\dHloc)=0.38 \kmsMpc$. Most of the increase
in this quantity comes from the skewed redshift distribution of $\nsnz$.  We
have checked that sampling 217 haloes randomly at $0.023<z<0.15$ barely
increases the scatter (0.13 $\kmsMpc$), while sampling 217 haloes according to 
$\nsnz$ gives the same results as $\nsnz/\nhaloz$ weighting (0.39 $\kmsMpc$).

The penultimate column of Table \ref{tab:dH0} shows the results when the rotations
of the SN coordinate system with respect to the simulation are taken into
account, without weighting (using equation~\ref{eq:dH0_averaged}); the
standard deviation rises slightly.  The last column is similar to the previous
one, with each SN weighted by the inverse of the square of the magnitude error
(using equation~\ref{eq:dH0_weighted}).   This weighting decreases the scatter
slightly because the magnitude error is smaller (thus weight is larger) for
high-redshift SNe.  This last column contains our best estimate for the sample
variance in $\dHloc$: approximately Gaussian distribution with
$\sigma(\dHloc)=0.31\kmsMpc$. We note that this value is similar to the 0.4\%
uncertainty due to sample variance quoted in R16 (see their fig.~12).

\section{Impact of local density contrast}\label{sec:delta}

\begin{figure*}
\includegraphics[width=\columnwidth]{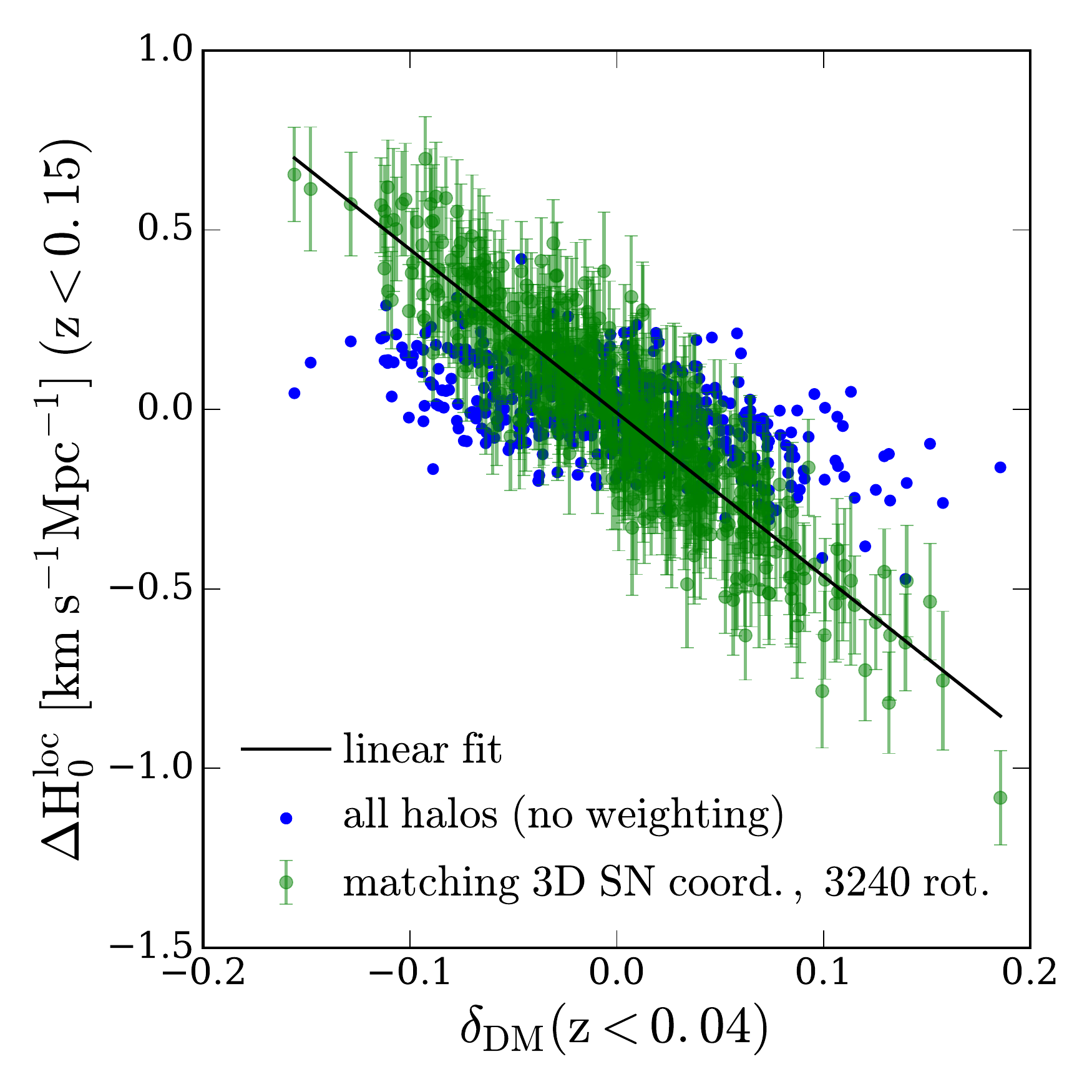}
\includegraphics[width=\columnwidth]{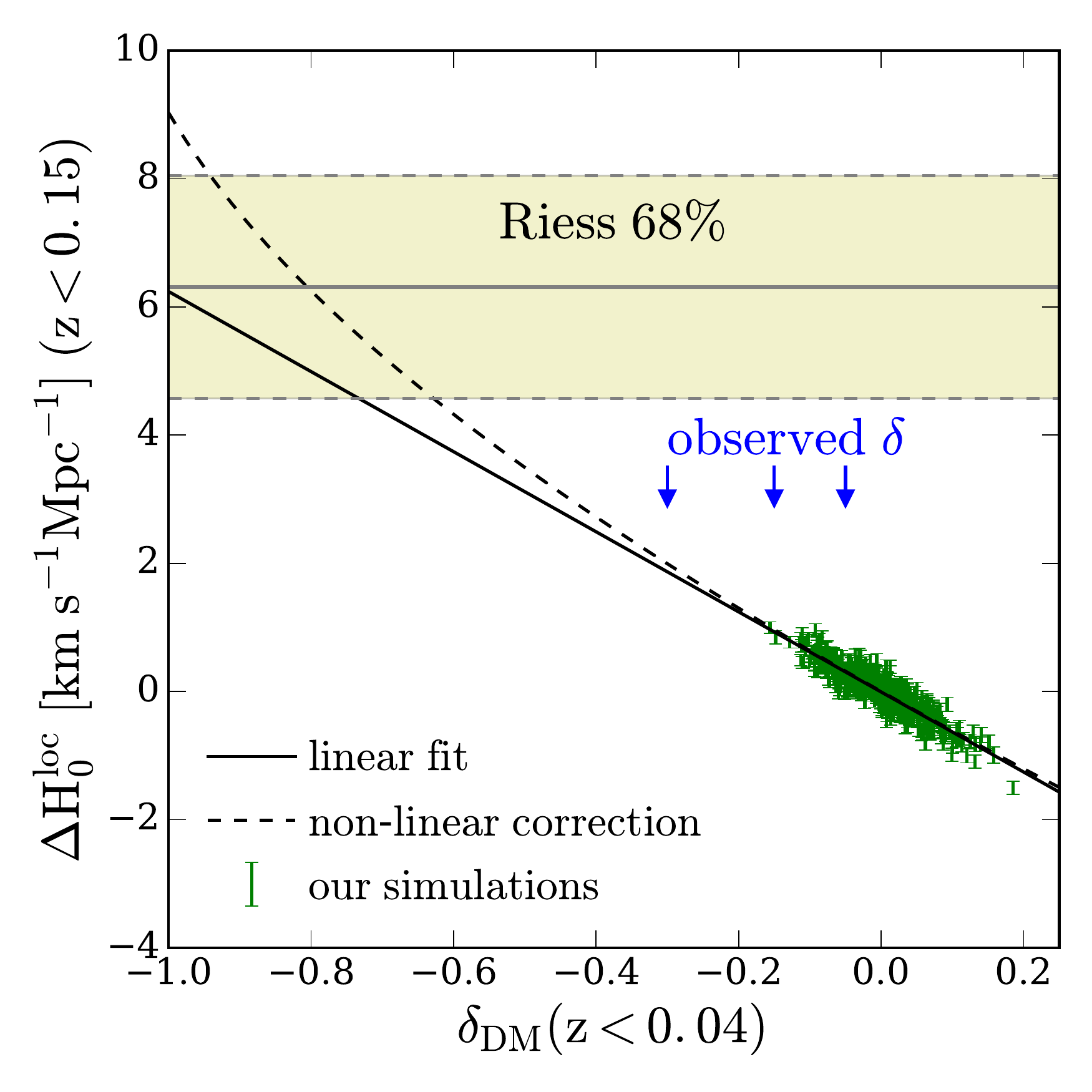}
\caption[]{Correlation between $\dHloc$ for $\zmax=0.15$ (corresponding
to the SN sample) and dark matter density contrast $\delta$ for $\zmax=0.04$
(corresponding to the distance scale for local density measurements); both are measured from 512
sub-volumes of the Dark Sky simulations.  \textit{Left}: $\dHloc$
measurements from matching the 3D coordinates of SNe and haloes in 
sub-volumes (green points with error bars), compared to
inference from all haloes in sub-volumes (blue points). The error bars on
the green points reflect the variances from rotations of the SN coordinate system
within each sub-volume. The solid line shows the linear fit to the green points.  
\textit{Right}: zoomed-out version of the left-hand panel.
We additionally mark the location of several $\delta$ values from observations,
as well as the 1-$\sigma$ range favoured by the R16 analysis.  We note that
none of the observations of $\delta$ can account for the 6 $\kmsMpc$
difference between $\Hloc$ and $\HCMB$.}
\label{fig:dHloc_delta_obs}
\end{figure*}

We now investigate the relation between fluctuations in $\Hloc$ and the
density contrast $\delta = (\rho - \bar{\rho}) / \bar{\rho}$, which can also be measured in the simulation. 
A negative/positive density contrast implies a higher/lower local expansion rate than the average value.
The deviation in $\Hloc$ is directly related to the local density contrast $\delta$
via
\beq
\frac{\dHloc}{\Hloc} = -\frac{1}{3} \delta f(\Omega_M) \Theta(\delta, \Omega_M) \ ,
\label{eq:dHloc}
\eeq
where $\Theta$ is the non-linear correction (\citealt{Marra13}), 
\beq
\Theta \simeq 1 - 0.0882 \delta - \frac{0.123 \sin\delta}{1.29 + \delta}    \ .
\label{eq:Theta}
\eeq
In a flat $\Lambda$CDM universe, 
\beq
f(\Omega_M,z) \approx \bigg(\Omega_M(z)\bigg)^\gamma \ ,  
\eeq
where $\gamma \approx 0.55$.  For the cosmology adopted by Dark Sky simulations, $f\approx 0.5$ at $z=0$.

In this section, we put the $\dHloc$ and $\delta$ measured from simulations in the context of several observational results.

\subsection{Observed local density contrast}\label{sec:obs_delta}

We consider three observations of the density in the local Universe.

\begin{itemize}

\item \cite{Keenan13} used the $K$-band (near-infrared) luminosity density to trace the matter density for $0.01 < z < 0.2$ over $600 \deg^2$ of the sky.  They found that the nearby Universe is underdense out to $\simeq$ 300 Mpc. Specifically, at $\simeq$ 200 Mpc, the density is $\simeq 30\%$ lower than at $\simeq$ 600 Mpc ($\delta=-0.3$, see their fig.~11).

\item \cite{WhitbournShanks14} used the $K$-band and $r$-band galaxy number densities out to $\simeq 300 \hiMpc$.  They included almost full-sky data but the depth is not uniform.  They found $\simeq$ 5, 15, and 40 \% underdensity ($\delta = -0.05$, $-0.15$, and $-0.4$) out to 150 $\hiMpc$, depending on the survey field (see their figs 3, 8 and 10). 

\item \cite{Carrick15} used the galaxy luminosity density from 2M++ to reconstruct the density of galaxies out to $200\hiMpc$.   They included full-sky results out to $\simeq 120\hiMpc$.
They found no evidence for local underdensity ($\delta = 0$); in fact, their results suggested slight overdensity (see their fig.~10).\footnote{Their luminosity-weighted density contrast and peculiar velocity are publicly available at \href{http://cosmicflows.iap.fr}{http://cosmicflows.iap.fr}.}

\end{itemize}

We emphasize that there are substantial sources of systematic uncertainty in
mapping out the mass density contrast from the luminosity and number density
of galaxies. Hence, we only include the observational constraints above as a
rough guideline on what the current data indicate.

\subsection{Comparing simulations and observations}

We next turn to the relation between density contrast and local measurements of
the Hubble constant in the simulations. We first verify that, as
expected, $\dHloc$ and $\delta$ measured over the same volume obey the linear-theory 
relation in equation (\ref{eq:dHloc}) if all haloes in the volume are
used in the $\Hloc$ measurement (as long as very nearby objects are excised to
avoid nonlinear effects).  In practice, however, the  observational
constraints of $\delta$ are limited to relatively low redshift (e.g.,
$\lesssim 120\hiMpc$ or $z \lesssim 0.04$ for \citealt{Carrick15}), while the
SN sample extends to $z=0.15$.

Despite this mismatch in redshift range, we have found that $\dHloc$
measured for $0.023 < z < 0.15$ (with the R16 selection) correlates strongly
with $\delta$ measured for $z<0.04$. This is unsurprising because the R16 SN
selection peaks at $z<0.04$, and thus $\dHloc$ measured for $z<0.15$ and that
for $z<0.04$ correlate strongly with each other. This correlation indicates
that we can infer $\dHloc$ from the $\delta$ measured at relatively low
redshift.

The left-hand panel of Fig.~\ref{fig:dHloc_delta_obs} shows the correlation between 
$\delta$ and $\dHloc$.  The blue points show the $\dHloc$ measured using all
haloes in sub-volumes, while the green points with error bars show  $\dHloc$ from 3240 rotations of each sub-volume.  Note that the
comparison to the standard theory expectation for the $\dHloc$--$\delta$
relation is not applicable here, since the two quantities are evaluated at
different redshifts. We fit a straight line to the points and extend it to low $\delta$.

The right-hand panel of Fig.~\ref{fig:dHloc_delta_obs} is the zoomed-out version of
the left-hand panel, with some additional information.  The green points with
errors again show the measurements from our simulation. The solid line is the
linear fit, and the dash curve has the non-linear correction (equation
\ref{eq:Theta}) applied to the linear fit.  The three vertical arrows denote the observational estimates
of $\delta= -0.3$, $-0.15$, and $-0.05$, outlined in
Section~\ref{sec:obs_delta}.  As can be seen, $\delta < -0.3$ is expected to
be extremely rare, as we do not find one in 512 sub-volumes.  Moreover, even
such a low $\delta$ could not account for the 6$\kmsMpc$ difference between
R16 and P16; the 6$\kmsMpc$ difference corresponds to $\delta\simeq
-0.8$.

We conclude that, while $\dHloc$ follows the expected trend with the local
mass density contrast $\delta$, the underdensity required to ameliorate the
discrepancy between the local and global measurements of $H_0$ far exceeds the
values of $\delta$ seen in simulations or in direct observations.

\section{Discussion}\label{sec:discussions}

We now discuss whether there are sources of sample variance 
not captured in our simulations that could significantly increase the sample
variance estimated in  Section~\ref{sec:results}
($\sigma(\dHloc)\simeq 0.31\kmsMpc$ for the R16 SN selection).

The first question is whether we have sufficient statistics in our
simulation. The total simulation volume, $(8\hiGpc)^3$, captures the effect
of sufficiently large modes and is subject to negligible variance due to modes
larger than the simulation volume.  In addition, we explicitly checked that our results
(e.g.,\ the total histogram in Fig.~\ref{fig:dH0_histograms}) are well
converged with respect to the number of sub-volumes, and our choice of 512
sub-volumes is  sufficient. This also guarantees that choosing the halo closest
to the centre of each sub-volume   as the location of the observer is
sufficient; varying the observer's location prescription within sub-volumes
would produce additional volume samples, which are unnecessary given that we
already have sufficient statistics. Obtaining more sub-volumes populated by SN,
either by choosing alternate locations of the observer in existing sub-volumes
or by simply using additional sub-volumes from, e.g., additional runs of Dark
Sky simulations, would help populate the far tails of the distribution of the
values of $\Hloc$, and produce a quantitative answer to the extremum-value
likelihood of the Hubble constant being greater than some threshold. However,
these additional runs would not change anything in our conclusion that sample
variance alone falls far short of reconciling the values of the global and
local measurements of $H_0$. In summary, we have sufficiently many sub-volumes
for simulating the $\Hloc$ measurements.

Next, we discuss the impact of velocity bias.  In our analysis, we assume that
the velocity of an SN is equal to the velocity of its host halo, 
and that the velocity of a halo traces the velocity field of the surrounding dark matter.
The former assumption holds for Milky Way mass haloes;
even if an SN has the maximum circular velocity of its host halo, 
its motion within the halo is negligible compared 
with the peculiar velocity of the host halo.
The latter assumption may break down due to dynamic friction or tidal stripping
\citep[see, e.g.,][]{Wu13bv}; in this work, we explicitly select isolated
haloes (haloes that are not within the virial radius of other haloes), and
therefore we expect the effect of velocity bias to be negligible.  Moreover,
a $\sim 5\%$ velocity bias is unlikely to be coherent over scales of hundreds
of Mpc, which is required in order to increase the locally inferred value of
$\Hloc$.

Another issue of interest is the choice of the mass of host haloes. In this
work, we use haloes with $\Mvir\in[10^{12.3}, 10^{12.4}]\,\Msun$ 
as SN hosts.  The SN host stellar mass information is partially available for the
Supercal sample \citep{Scolnic14}; it has a fairly wide distribution that
peaks around $M_\mathrm{stellar}\simeq 10^{11}\Msun$.  Given that
$\Mvir/M_\mathrm{stellar}\sim O(100)$ for this mass range \citep[see,
e.g.,][]{Behroozi13},  our assumption that the SN host haloes lie in the range
$\Mvir\in[10^{12.3}, 10^{12.4}]\,\Msun$ is reasonable. We have nevertheless
verified that we get essentially the same results if we match the SNe to all
haloes above $10^{13}\Msun$.

On the whole, we conclude that our analysis is robust with respect to the
statistics of simulated volumes  and the halo mass selection.
More specifically,  our assignment of haloes to SN hosts is faithful to
observations, and we do not expect that other choices for the assignment
--- or in any other step of our calculation --- would
substantially change our quantitative conclusions about the size of the sample
variance in $\Hloc$ measurements.

\section{Conclusions}\label{sec:conclusions}

In this paper, we have studied the sample variance in the local measurements
of the Hubble constant. A statistical fluctuation in this measurement, due to
the fact that the volume probed is a small fraction of the volume enclosed by
the surface of last scattering, is arguably the simplest explanation why the
local measurements, $\Hloc$, would be biased relative to the global value
inferred from the CMB, $\HCMB$. We have been particularly motivated by the
possibility that the spatially inhomogeneous distribution of SNe, from which the
recession velocities and thus the Hubble constant are inferred, could further
increase the sample variance on top of what has been estimated in the
literature. To account for the SN selection, we, for the first time,
explicitly model the exact, inhomogeneous three-dimensional distribution of SNe
used in the $\Hloc$ measurements, and we comb through more than one million
possible ways in which these SNe could sample the velocity field given by
haloes in the 8$\hiGpc$ $N$-body Dark Sky simulations.

For observers centred on haloes of virial mass $\sim10^{12.3}\Msun$, we find
that the measurement of $\Hloc$ is on average unbiased for all scales, while
the statistical error depends on the redshift distribution of SNe. Employing
the skewed redshift distribution and sparse angular distribution of SNe
utilized in R16, we find a final statistical error of $\sigma(\dHloc) = 0.31
\kmsMpc$, much smaller than the $\sim 6\kmsMpc$ discrepancy between the local
and global $H_0$ measurements. Our results are robust with respect to the halo
mass range chosen.

We next turn to the relation between the local measurements of the Hubble constant
and density contrast. As expected, $\dHloc$ and $\delta$ measured over the same
volume obey the relation from linear perturbation theory if all haloes
in the volume are used in the $\dHloc$ measurement. However, once the SN-matched 
haloes are used to mimic the actual measurement, the resulting non-uniform 
sampling of the volume (and hence the velocity field) spoils the
expected $\dHloc-\delta$ relation. Nevertheless, we find a linear correlation
between $\dHloc (\zmax=0.15)$ and density measured in a much smaller volume
effectively sampled by SN, $\delta(\zmax=0.04)$. We find that resolving the
$H_0$ measurement discrepancy by postulating a local void requires an
underdensity of $\delta \simeq -0.8$ with a radius of $120 \hiMpc$, which is
extremely unlikely in a $\Lambda$CDM universe.  Existing direct observational
constraints on the mass density at this scale, although highly uncertain,
also exclude such a low density.

It is worth noting that R16 explicitly corrected for the peculiar
velocity of each SN using a reconstruction of the local density field. This
correction, if accurate, explicitly removes the effects of the local
over/underdensity, and hence sample variance. Density field
reconstructions are subject to a range of systematic errors, and therefore the sample
variance correction is necessarily expected to be only partial. 
Overall, our results are conservative: the sample variance in the measured $\Hloc$ is
{\it at most} the value found in our analysis, and likely just a fraction
thereof due to explicit peculiar velocity corrections.

In summary, assuming the standard $\Lambda$CDM cosmological model, the current
tension between $\Hloc$ and $\HCMB$ is unlikely to be resolved or even
appreciably ameliorated by invoking sample variance. Given that the tension
between the two measurements cannot be fully attributed to the sample variance
of the local measurements, is imperative to understand the systematics in
local and CMB experiments, as well as the possible new physics behind the tension.

\section*{Acknowledgments}

We thank Adam Riess, Dan Shafer, and Radek Wojtak for insightful suggestions, 
Matt Turk for assistance for accessing the Dark Sky simulations, and Dan Scolnic for
providing the Supercal data set.  H.W.\ acknowledges the support by the U.S.\
National Science Foundation (NSF) grant AST1313037. D.H.\ is supported by NSF
under contract AST-0807564 and the Department of Energy under contract DE-FG02-95ER40899.  
The calculations in this work were performed on the Caltech
computer cluster Zwicky, which is supported by NSF MRI-R2 award number
PHY-096029, and on {\tt kingjames}, lovingly supported by his majesty's owner
at the University of Michigan.

\appendix
\section{Impact of maximum redshift and weighting}\label{app:weighting}
\begin{figure}
\includegraphics[width=\columnwidth]{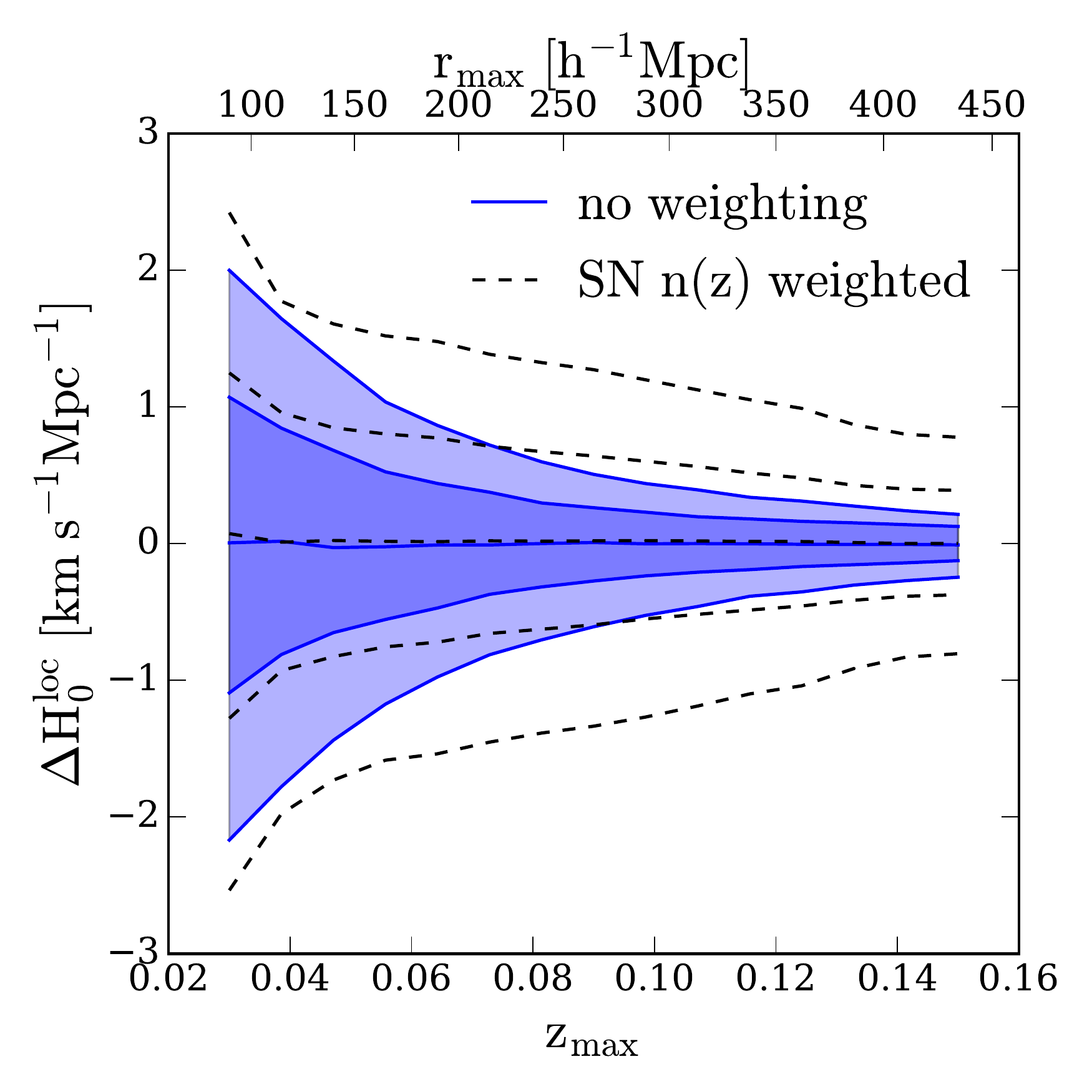}
\caption[]{Deviations of the locally-measured Hubble constant
  as a function of maximum redshift (bottom $x$-axis) and distance
  (top $x$-axis).  Observers and SNe are centred on haloes within $10^{12.3} <
  \Mvir/\Msun < 10^{12.4}$.  The blue curves and shades correspond to no
  weighting, and the black dash curves correspond to weighting with the
  redshift distribution of SNe in R16.  The curves correspond to the median and
  the 68\% and 95\% intervals.  We see that, at this halo mass, $\Hloc$ is
  essentially unbiased on all scales.  Using the realistic redshift
  distribution leads to a larger statistical error.}
\label{fig:dHloc_zmax}
\end{figure}

In this appendix, we discuss the weighting of haloes based on $\nsnz/\nhaloz$.
We first demonstrate that, as mentioned in Section~\ref{sec:darksky},  our
choice of halo mass, $[10^{12.3}, 10^{12.4}]\Msun$, leads to unbiased $\Hloc$
at all scales.  Fig.~\ref{fig:dHloc_zmax} shows how the deviation of the locally measured
Hubble constant ($\dHloc$) depends on the maximum redshift and distance of the
SN sample.  The blue curves and shades correspond to using all haloes between
$\zmin = 0.023$ and $\zmax$ (varies along the $x$-axis), and different curves
correspond to the median and the 68\% and 95\% intervals.  Clearly, when observers
are located on haloes of this mass, the median of $\Hloc$ is unbiased for all
scales because haloes of this mass occupy approximately unbiased environments. The
statistical error of $\Hloc$ drops precipitously with redshift due to the
rapidly increasing number of haloes.  We note that W14 found $\dHloc < 0 $
because they put observers on more massive haloes, which tend to be in overdense environments.

The black dash curves in Fig.~\ref{fig:dHloc_zmax} show  how $\dHloc$
depends on $\zmax$ when we weight haloes by $\nsnz/\nhaloz$. We see that the
median remains unbiased, while the statistical error becomes significantly
larger than the case of all haloes.  The statistical errors decrease
slowly with redshift, reflecting the rarity of SN at higher redshift.  For
$\zmax=0.15$, the 68\% statistical error is 0.5\%.

\section{Estimators of $\dHloc$}\label{app:estimators}
\begin{figure}
\includegraphics[width=\columnwidth]{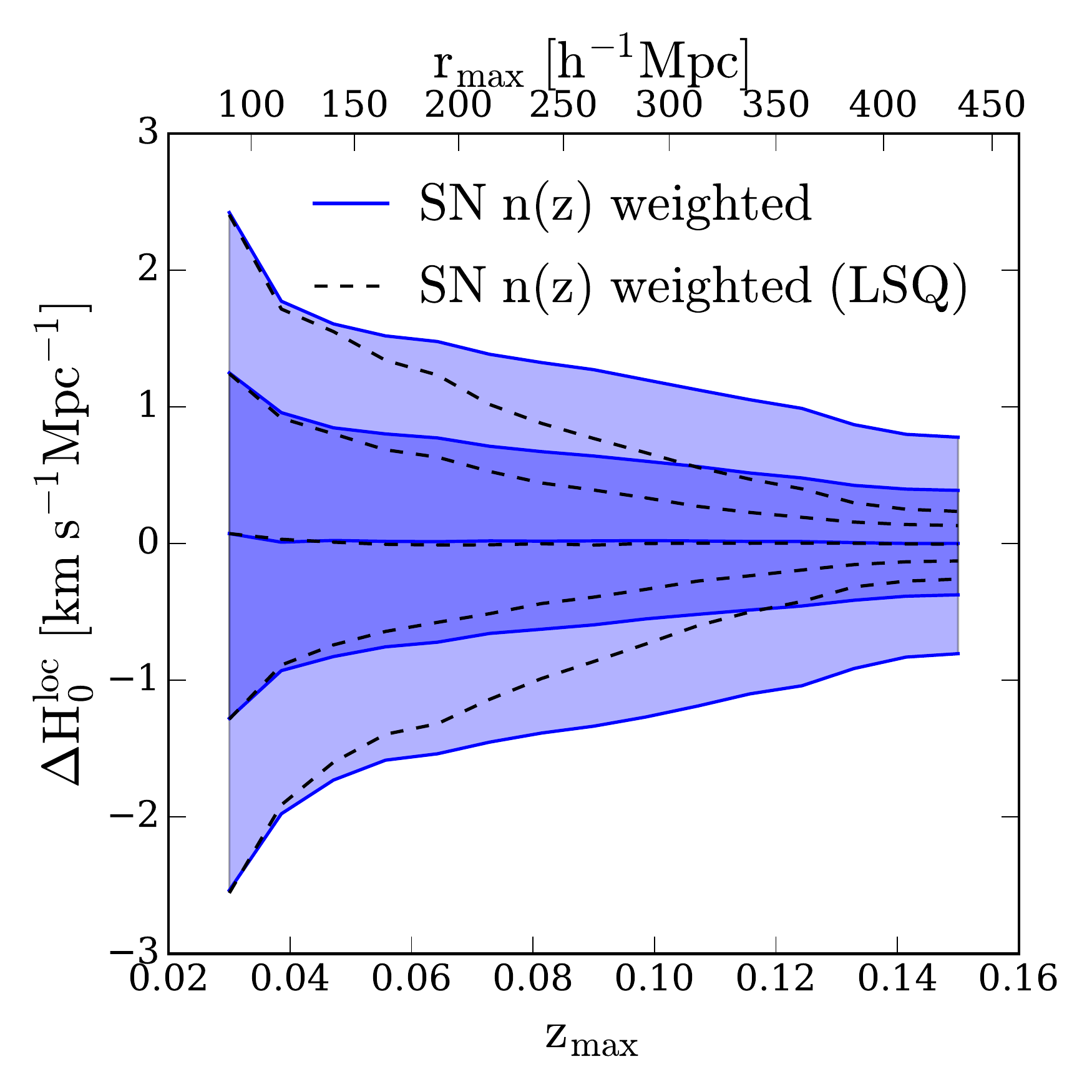}
\caption[]{Comparison between our estimator (equation~\ref{eq:dHloc_ours}) and the
  LSQ estimator (equation~\ref{eq:dHloc_LSQ}), weighted with the redshift
  distribution of SNe in R16.  As can be seen, the LSQ estimator systematically
  gives smaller $\dHloc$, and the difference is larger for larger $\zmax$.}
\label{fig:estimators}
\end{figure}
In Section~\ref{sec:dHloc}, we have derived the estimator for $\dHloc$, 
\beq
\dHloc  = \frac{\sum w_i {\bf v_i}\cdot {\bf r_i}/r_i^2 }{\sum w_i } \ ,
\label{eq:dHloc_ours}
\eeq
which is relevant for the SN analysis in R16.  
In this appendix, we discuss the LSQ estimator, which is sometimes seen in the literature.

For $z \ll 1$, the peculiar velocity affects the local measurements of the Hubble constant $\Hloc$ via 
\beq
H_0 r + v_r = \Hloc r  \ .
\eeq
Let us assume that the $i^{th}$ SN in our sample has a position vector 
${\bf r_i}$ with respect to the Milky Way, a peculiar velocity ${\bf v_i}$ with respect to the CMB rest frame, and weighting $w_i$.  
The problem is equivalent to a weighted linear regression with zero intercept; thus, $\dHloc$ is given by
the LSQ estimator
\beq
\dHloc = \Hloc - H_0 = \frac{\sum w_i {\bf v_i}\cdot {\bf r_i} }{\sum w_i r_i^2} \quad\mbox{(LSQ)} \ .
\label{eq:dHloc_LSQ}
\eeq
While the LSQ estimator is the correct one for estimating the slope of a
relation (i.e., using $v_r$ and $r$ to estimate $\dHloc$), it does not apply
to SN measurements, given that $v_r$ and $r$ are not separately
available. Rather, SNe provide measurements of the specific combination
$v_r/r$, which corresponds to the deviation from the global Hubble
constant, and hence the appropriate estimator is given in equation
(\ref{eq:dHloc_ours}).

Fig.~\ref{fig:estimators} shows the comparison between our estimator  and the
LSQ estimator for $\dHloc$.   We weight the haloes by $\nsnz/\nhaloz$ and show
$\dHloc$ as a function  of the maximum redshift of the survey.   Our estimator
produces a larger scatter of $\dHloc$ than the LSQ estimator,  and the
difference increases with redshift. At $\zmax = 0.15$, our estimator produces
approximately three times larger scatter than the LSQ estimator.

\section{Sampling different orientations: Euler angles}\label{app:Euler}

Here we provide more details about varying the orientation of the SN
coordinate system relative to the simulation coordinate
system. We apply Euler angles to uniformly rotate the coordinates of SNe.
Specifically, we use the ``$x$--$z$--$x$'' convention; that is, we first
rotate the coordinate system around the $x$-axis by $\alpha$, second rotate
around the $z'$-axis by $\beta$, and third rotate the $x''$-axis by $\gamma$.

The ranges of the angles are given by \beqa \alpha &\in [0, 2\pi)  \\
\cos\beta &\in [1,-1] \\ \gamma &\in [0, 2\pi)  \ . \eeqa For $\alpha$ and
$\gamma$, we use 18 angles spaced by 20$^\circ$ (0$^\circ$ to 340$^\circ$).
For $\beta$, we use 10 angles with equally spaced $\cos\beta$ (-1 to 1). This
leads to a total of $(360/20)^2(180/20+1)=3240$ possible orientations.

\bibliographystyle{mnras}
\bibliography{H0-CV}

\begin{thebibliography}{}
\makeatletter
\relax
\def\mn@urlcharsother{\let\do\@makeother \do\$\do\&\do\#\do\^\do\_\do\%\do\~}
\def\mn@doi{\begingroup\mn@urlcharsother \@ifnextchar [ {\mn@doi@}
  {\mn@doi@[]}}
\def\mn@doi@[#1]#2{\def\@tempa{#1}\ifx\@tempa\@empty \href
  {http://dx.doi.org/#2} {doi:#2}\else \href {http://dx.doi.org/#2} {#1}\fi
  \endgroup}
\def\mn@eprint#1#2{\mn@eprint@#1:#2::\@nil}
\def\mn@eprint@arXiv#1{\href {http://arxiv.org/abs/#1} {{\tt arXiv:#1}}}
\def\mn@eprint@dblp#1{\href {http://dblp.uni-trier.de/rec/bibtex/#1.xml}
  {dblp:#1}}
\def\mn@eprint@#1:#2:#3:#4\@nil{\def\@tempa {#1}\def\@tempb {#2}\def\@tempc
  {#3}\ifx \@tempc \@empty \let \@tempc \@tempb \let \@tempb \@tempa \fi \ifx
  \@tempb \@empty \def\@tempb {arXiv}\fi \@ifundefined
  {mn@eprint@\@tempb}{\@tempb:\@tempc}{\expandafter \expandafter \csname
  mn@eprint@\@tempb\endcsname \expandafter{\@tempc}}}

\bibitem[\protect\citeauthoryear{{Addison}, {Huang}, {Watts}, {Bennett},
  {Halpern}, {Hinshaw}  \& {Weiland}}{{Addison} et~al.}{2016}]{Addison16}
{Addison} G.~E.,  {Huang} Y.,  {Watts} D.~J.,  {Bennett} C.~L.,  {Halpern} M.,
  {Hinshaw} G.,   {Weiland} J.~L.,  2016, \mn@doi [\apj]
  {10.3847/0004-637X/818/2/132}, \href
  {http://adsabs.harvard.edu/abs/2016ApJ...818..132A} {818, 132}

\bibitem[\protect\citeauthoryear{{Aubourg} et~al.,}{{Aubourg}
  et~al.}{2015}]{Aubourg15}
{Aubourg} {\'E}.,  et~al., 2015, \mn@doi [\prd] {10.1103/PhysRevD.92.123516},
  \href {http://adsabs.harvard.edu/abs/2015PhRvD..92l3516A} {92, 123516}

\bibitem[\protect\citeauthoryear{{Behroozi}, {Wechsler}  \& {Wu}}{{Behroozi}
  et~al.}{2013a}]{Behroozi13rs}
{Behroozi} P.~S.,  {Wechsler} R.~H.,   {Wu} H.-Y.,  2013a, \mn@doi [\apj]
  {10.1088/0004-637X/762/2/109}, \href
  {http://adsabs.harvard.edu/abs/2013ApJ...762..109B} {762, 109}

\bibitem[\protect\citeauthoryear{{Behroozi}, {Wechsler}  \&
  {Conroy}}{{Behroozi} et~al.}{2013b}]{Behroozi13}
{Behroozi} P.~S.,  {Wechsler} R.~H.,   {Conroy} C.,  2013b, \mn@doi [\apj]
  {10.1088/0004-637X/770/1/57}, \href
  {http://adsabs.harvard.edu/abs/2013ApJ...770...57B} {770, 57}

\bibitem[\protect\citeauthoryear{{Ben-Dayan}, {Durrer}, {Marozzi}  \&
  {Schwarz}}{{Ben-Dayan} et~al.}{2014}]{BenDayan14}
{Ben-Dayan} I.,  {Durrer} R.,  {Marozzi} G.,   {Schwarz} D.~J.,  2014, \mn@doi
  [Physical Review Letters] {10.1103/PhysRevLett.112.221301}, \href
  {http://adsabs.harvard.edu/abs/2014PhRvL.112v1301B} {112, 221301}

\bibitem[\protect\citeauthoryear{{Bland-Hawthorn} \&
  {Gerhard}}{{Bland-Hawthorn} \& {Gerhard}}{2016}]{Bland-Hawthorn16}
{Bland-Hawthorn} J.,  {Gerhard} O.,  2016, \mn@doi [\araa]
  {10.1146/annurev-astro-081915-023441}, \href
  {http://adsabs.harvard.edu/abs/2016ARA%26A..54..529B} {54, 529}

\bibitem[\protect\citeauthoryear{{Bonvin} et~al.,}{{Bonvin}
  et~al.}{2017}]{Bonvin17}
{Bonvin} V.,  et~al., 2017, \mn@doi [\mnras] {10.1093/mnras/stw3006}, \href
  {http://adsabs.harvard.edu/abs/2017MNRAS.465.4914B} {465, 4914}

\bibitem[\protect\citeauthoryear{{Cardona}, {Kunz}  \& {Pettorino}}{{Cardona}
  et~al.}{2017}]{Cardona17}
{Cardona} W.,  {Kunz} M.,   {Pettorino} V.,  2017, \mn@doi [\jcap]
  {10.1088/1475-7516/2017/03/056}, \href
  {http://adsabs.harvard.edu/abs/2017JCAP...03..056C} {3, 056}

\bibitem[\protect\citeauthoryear{{Carrick}, {Turnbull}, {Lavaux}  \&
  {Hudson}}{{Carrick} et~al.}{2015}]{Carrick15}
{Carrick} J.,  {Turnbull} S.~J.,  {Lavaux} G.,   {Hudson} M.~J.,  2015, \mn@doi
  [\mnras] {10.1093/mnras/stv547}, \href
  {http://adsabs.harvard.edu/abs/2015MNRAS.450..317C} {450, 317}

\bibitem[\protect\citeauthoryear{{Cooray} \& {Caldwell}}{{Cooray} \&
  {Caldwell}}{2006}]{CoorayCaldwell06}
{Cooray} A.,  {Caldwell} R.~R.,  2006, \mn@doi [\prd]
  {10.1103/PhysRevD.73.103002}, \href
  {http://adsabs.harvard.edu/abs/2006PhRvD..73j3002C} {73, 103002}

\bibitem[\protect\citeauthoryear{{Courtois}, {Pomar{\`e}de}, {Tully}, {Hoffman}
   \& {Courtois}}{{Courtois} et~al.}{2013}]{Courtois13}
{Courtois} H.~M.,  {Pomar{\`e}de} D.,  {Tully} R.~B.,  {Hoffman} Y.,
  {Courtois} D.,  2013, \mn@doi [\aj] {10.1088/0004-6256/146/3/69}, \href
  {http://adsabs.harvard.edu/abs/2013AJ....146...69C} {146, 69}

\bibitem[\protect\citeauthoryear{{Efstathiou}}{{Efstathiou}}{2014}]{Efstathiou14}
{Efstathiou} G.,  2014, \mn@doi [\mnras] {10.1093/mnras/stu278}, \href
  {http://adsabs.harvard.edu/abs/2014MNRAS.440.1138E} {440, 1138}

\bibitem[\protect\citeauthoryear{{Feeney}, {Mortlock}  \& {Dalmasso}}{{Feeney}
  et~al.}{2017}]{Feeney17}
{Feeney} S.~M.,  {Mortlock} D.~J.,   {Dalmasso} N.,  2017, preprint, \href
  {http://adsabs.harvard.edu/abs/2017arXiv170700007F} {} (\mn@eprint {arXiv}
  {1707.00007})

\bibitem[\protect\citeauthoryear{{Fleury}, {Clarkson}  \& {Maartens}}{{Fleury}
  et~al.}{2017}]{Fleury17}
{Fleury} P.,  {Clarkson} C.,   {Maartens} R.,  2017, \mn@doi [\jcap]
  {10.1088/1475-7516/2017/03/062}, \href
  {http://adsabs.harvard.edu/abs/2017JCAP...03..062F} {3, 062}

\bibitem[\protect\citeauthoryear{{Follin} \& {Knox}}{{Follin} \&
  {Knox}}{2017}]{Follin17}
{Follin} B.,  {Knox} L.,  2017, preprint, \href
  {http://adsabs.harvard.edu/abs/2017arXiv170701175F} {} (\mn@eprint {arXiv}
  {1707.01175})

\bibitem[\protect\citeauthoryear{{Freedman} \& {Madore}}{{Freedman} \&
  {Madore}}{2010}]{Freedman10}
{Freedman} W.~L.,  {Madore} B.~F.,  2010, \mn@doi [\araa]
  {10.1146/annurev-astro-082708-101829}, \href
  {http://adsabs.harvard.edu/abs/2010ARA%26A..48..673F} {48, 673}

\bibitem[\protect\citeauthoryear{{Freedman} et~al.,}{{Freedman}
  et~al.}{2001}]{Freedman01}
{Freedman} W.~L.,  et~al., 2001, \mn@doi [\apj] {10.1086/320638}, \href
  {http://adsabs.harvard.edu/abs/2001ApJ...553...47F} {553, 47}

\bibitem[\protect\citeauthoryear{{Freedman}, {Madore}, {Scowcroft}, {Burns},
  {Monson}, {Persson}, {Seibert}  \& {Rigby}}{{Freedman}
  et~al.}{2012}]{Freedman12}
{Freedman} W.~L.,  {Madore} B.~F.,  {Scowcroft} V.,  {Burns} C.,  {Monson} A.,
  {Persson} S.~E.,  {Seibert} M.,   {Rigby} J.,  2012, \mn@doi [\apj]
  {10.1088/0004-637X/758/1/24}, \href
  {http://adsabs.harvard.edu/abs/2012ApJ...758...24F} {758, 24}

\bibitem[\protect\citeauthoryear{{Guy} et~al.,}{{Guy} et~al.}{2010}]{Guy10}
{Guy} J.,  et~al., 2010, \mn@doi [\aap] {10.1051/0004-6361/201014468}, \href
  {http://adsabs.harvard.edu/abs/2010A%26A...523A...7G} {523, A7}

\bibitem[\protect\citeauthoryear{{Hinshaw} et~al.,}{{Hinshaw}
  et~al.}{2013}]{WMAP9}
{Hinshaw} G.,  et~al., 2013, \mn@doi [\apjs] {10.1088/0067-0049/208/2/19},
  \href {http://adsabs.harvard.edu/abs/2013ApJS..208...19H} {208, 19}

\bibitem[\protect\citeauthoryear{{Hubble}}{{Hubble}}{1929}]{Hubble29}
{Hubble} E.,  1929, \mn@doi [Proceedings of the National Academy of Science]
  {10.1073/pnas.15.3.168}, \href
  {http://adsabs.harvard.edu/abs/1929PNAS...15..168H} {15, 168}

\bibitem[\protect\citeauthoryear{{Hui} \& {Greene}}{{Hui} \&
  {Greene}}{2006}]{HuiGreene06}
{Hui} L.,  {Greene} P.~B.,  2006, \mn@doi [\prd] {10.1103/PhysRevD.73.123526},
  \href {http://adsabs.harvard.edu/abs/2006PhRvD..73l3526H} {73, 123526}

\bibitem[\protect\citeauthoryear{{Jha}, {Riess}  \& {Kirshner}}{{Jha}
  et~al.}{2007}]{Jha07}
{Jha} S.,  {Riess} A.~G.,   {Kirshner} R.~P.,  2007, \mn@doi [\apj]
  {10.1086/512054}, \href {http://adsabs.harvard.edu/abs/2007ApJ...659..122J}
  {659, 122}

\bibitem[\protect\citeauthoryear{{Keenan}, {Barger}  \& {Cowie}}{{Keenan}
  et~al.}{2013}]{Keenan13}
{Keenan} R.~C.,  {Barger} A.~J.,   {Cowie} L.~L.,  2013, \mn@doi [\apj]
  {10.1088/0004-637X/775/1/62}, \href
  {http://adsabs.harvard.edu/abs/2013ApJ...775...62K} {775, 62}

\bibitem[\protect\citeauthoryear{{Marra}, {Amendola}, {Sawicki}  \&
  {Valkenburg}}{{Marra} et~al.}{2013}]{Marra13}
{Marra} V.,  {Amendola} L.,  {Sawicki} I.,   {Valkenburg} W.,  2013, \mn@doi
  [Physical Review Letters] {10.1103/PhysRevLett.110.241305}, \href
  {http://adsabs.harvard.edu/abs/2013PhRvL.110x1305M} {110, 241305}

\bibitem[\protect\citeauthoryear{{Martinez-Vaquero}, {Yepes}, {Hoffman},
  {Gottl{\"o}ber}  \& {Sivan}}{{Martinez-Vaquero}
  et~al.}{2009}]{Martinez-Vaquero09}
{Martinez-Vaquero} L.~A.,  {Yepes} G.,  {Hoffman} Y.,  {Gottl{\"o}ber} S.,
  {Sivan} M.,  2009, \mn@doi [\mnras] {10.1111/j.1365-2966.2009.15093.x}, \href
  {http://adsabs.harvard.edu/abs/2009MNRAS.397.2070M} {397, 2070}

\bibitem[\protect\citeauthoryear{{Odderskov}, {Hannestad}  \&
  {Haugb{\o}lle}}{{Odderskov} et~al.}{2014}]{Odderskov14}
{Odderskov} I.,  {Hannestad} S.,   {Haugb{\o}lle} T.,  2014, \mn@doi [\jcap]
  {10.1088/1475-7516/2014/10/028}, \href
  {http://adsabs.harvard.edu/abs/2014JCAP...10..028O} {10, 028}

\bibitem[\protect\citeauthoryear{{Odderskov}, {Hannestad}  \&
  {Brandbyge}}{{Odderskov} et~al.}{2017}]{Odderskov17}
{Odderskov} I.,  {Hannestad} S.,   {Brandbyge} J.,  2017, \mn@doi [\jcap]
  {10.1088/1475-7516/2017/03/022}, \href
  {http://adsabs.harvard.edu/abs/2017JCAP...03..022O} {3, 022}

\bibitem[\protect\citeauthoryear{{Planck Collaboration Int. XLVI.}}{{Planck
  Collaboration Int. XLVI.}}{2016}]{Planck1605.02985}
{Planck Collaboration Int. XLVI.} 2016, \mn@doi [\aap]
  {10.1051/0004-6361/201628890}, \href
  {http://adsabs.harvard.edu/abs/2016A%26A...596A.107P} {596, A107}

\bibitem[\protect\citeauthoryear{{Planck Collaboration XIII.}}{{Planck
  Collaboration XIII.}}{2016}]{Planck15Cosmo}
{Planck Collaboration XIII.} 2016, \mn@doi [\aap]
  {10.1051/0004-6361/201525830}, \href
  {http://adsabs.harvard.edu/abs/2016A%26A...594A..13P} {594, A13}

\bibitem[\protect\citeauthoryear{{Planck Collaboration XVI.}}{{Planck
  Collaboration XVI.}}{2014}]{Planck13Cosmo}
{Planck Collaboration XVI.} 2014, \mn@doi [\aap] {10.1051/0004-6361/201321591},
  \href {http://adsabs.harvard.edu/abs/2014A%26A...571A..16P} {571, A16}

\bibitem[\protect\citeauthoryear{{Riess} et~al.,}{{Riess}
  et~al.}{2009}]{Riess09}
{Riess} A.~G.,  et~al., 2009, \mn@doi [\apj] {10.1088/0004-637X/699/1/539},
  \href {http://adsabs.harvard.edu/abs/2009ApJ...699..539R} {699, 539}

\bibitem[\protect\citeauthoryear{{Riess} et~al.,}{{Riess}
  et~al.}{2011}]{Riess11}
{Riess} A.~G.,  et~al., 2011, \mn@doi [\apj] {10.1088/0004-637X/730/2/119},
  \href {http://adsabs.harvard.edu/abs/2011ApJ...730..119R} {730, 119}

\bibitem[\protect\citeauthoryear{{Riess} et~al.,}{{Riess}
  et~al.}{2016}]{Riess16}
{Riess} A.~G.,  et~al., 2016, \mn@doi [\apj] {10.3847/0004-637X/826/1/56},
  \href {http://adsabs.harvard.edu/abs/2016ApJ...826...56R} {826, 56}

\bibitem[\protect\citeauthoryear{{Sandage} \& {Tammann}}{{Sandage} \&
  {Tammann}}{1982}]{Sandage82}
{Sandage} A.,  {Tammann} G.~A.,  1982, \mn@doi [\apj] {10.1086/159911}, \href
  {http://adsabs.harvard.edu/abs/1982ApJ...256..339S} {256, 339}

\bibitem[\protect\citeauthoryear{{Scolnic} \& {Kessler}}{{Scolnic} \&
  {Kessler}}{2016}]{ScolnicKessler16}
{Scolnic} D.,  {Kessler} R.,  2016, \mn@doi [\apjl]
  {10.3847/2041-8205/822/2/L35}, \href
  {http://adsabs.harvard.edu/abs/2016ApJ...822L..35S} {822, L35}

\bibitem[\protect\citeauthoryear{{Scolnic} et~al.,}{{Scolnic}
  et~al.}{2014}]{Scolnic14}
{Scolnic} D.,  et~al., 2014, \mn@doi [\apj] {10.1088/0004-637X/795/1/45}, \href
  {http://adsabs.harvard.edu/abs/2014ApJ...795...45S} {795, 45}

\bibitem[\protect\citeauthoryear{{Scolnic} et~al.,}{{Scolnic}
  et~al.}{2015}]{Scolnic15}
{Scolnic} D.,  et~al., 2015, \mn@doi [\apj] {10.1088/0004-637X/815/2/117},
  \href {http://adsabs.harvard.edu/abs/2015ApJ...815..117S} {815, 117}

\bibitem[\protect\citeauthoryear{{Shi} \& {Turner}}{{Shi} \&
  {Turner}}{1998}]{Shi98}
{Shi} X.,  {Turner} M.~S.,  1998, \mn@doi [\apj] {10.1086/305169}, \href
  {http://adsabs.harvard.edu/abs/1998ApJ...493..519S} {493, 519}

\bibitem[\protect\citeauthoryear{{Sinclair}, {Davis}  \&
  {Haugb{\o}lle}}{{Sinclair} et~al.}{2010}]{Sinclair10}
{Sinclair} B.,  {Davis} T.~M.,   {Haugb{\o}lle} T.,  2010, \mn@doi [\apj]
  {10.1088/0004-637X/718/2/1445}, \href
  {http://adsabs.harvard.edu/abs/2010ApJ...718.1445S} {718, 1445}

\bibitem[\protect\citeauthoryear{{Skillman}, {Warren}, {Turk}, {Wechsler},
  {Holz}  \& {Sutter}}{{Skillman} et~al.}{2014}]{Skillman14}
{Skillman} S.~W.,  {Warren} M.~S.,  {Turk} M.~J.,  {Wechsler} R.~H.,  {Holz}
  D.~E.,   {Sutter} P.~M.,  2014, preprint, \href
  {http://adsabs.harvard.edu/abs/2014arXiv1407.2600S} {} (\mn@eprint {arXiv}
  {1407.2600})

\bibitem[\protect\citeauthoryear{{Sorce}, {Tully}  \& {Courtois}}{{Sorce}
  et~al.}{2012}]{Sorce12}
{Sorce} J.~G.,  {Tully} R.~B.,   {Courtois} H.~M.,  2012, \mn@doi [\apjl]
  {10.1088/2041-8205/758/1/L12}, \href
  {http://adsabs.harvard.edu/abs/2012ApJ...758L..12S} {758, L12}

\bibitem[\protect\citeauthoryear{{Suyu} et~al.,}{{Suyu} et~al.}{2013}]{Suyu13}
{Suyu} S.~H.,  et~al., 2013, \mn@doi [\apj] {10.1088/0004-637X/766/2/70}, \href
  {http://adsabs.harvard.edu/abs/2013ApJ...766...70S} {766, 70}

\bibitem[\protect\citeauthoryear{{Tammann} \& {Reindl}}{{Tammann} \&
  {Reindl}}{2013}]{Tammann13}
{Tammann} G.~A.,  {Reindl} B.,  2013, \mn@doi [\aap]
  {10.1051/0004-6361/201219671}, \href
  {http://adsabs.harvard.edu/abs/2013A%26A...549A.136T} {549, A136}

\bibitem[\protect\citeauthoryear{{Turk}, {Smith}, {Oishi}, {Skory}, {Skillman},
  {Abel}  \& {Norman}}{{Turk} et~al.}{2011}]{Turk11}
{Turk} M.~J.,  {Smith} B.~D.,  {Oishi} J.~S.,  {Skory} S.,  {Skillman} S.~W.,
  {Abel} T.,   {Norman} M.~L.,  2011, \mn@doi [\apjs]
  {10.1088/0067-0049/192/1/9}, \href
  {http://adsabs.harvard.edu/abs/2011ApJS..192....9T} {192, 9}

\bibitem[\protect\citeauthoryear{{Turner}, {Cen}  \& {Ostriker}}{{Turner}
  et~al.}{1992}]{Turner92}
{Turner} E.~L.,  {Cen} R.,   {Ostriker} J.~P.,  1992, \mn@doi [\aj]
  {10.1086/116156}, \href {http://adsabs.harvard.edu/abs/1992AJ....103.1427T}
  {103, 1427}

\bibitem[\protect\citeauthoryear{{Wang}, {Spergel}  \& {Turner}}{{Wang}
  et~al.}{1998}]{Wang98}
{Wang} Y.,  {Spergel} D.~N.,   {Turner} E.~L.,  1998, \mn@doi [\apj]
  {10.1086/305539}, \href {http://adsabs.harvard.edu/abs/1998ApJ...498....1W}
  {498, 1}

\bibitem[\protect\citeauthoryear{{Warren}}{{Warren}}{2013}]{Warren13}
{Warren} M.~S.,  2013, preprint, \href
  {http://adsabs.harvard.edu/abs/2013arXiv1310.4502W} {} (\mn@eprint {arXiv}
  {1310.4502})

\bibitem[\protect\citeauthoryear{{Whitbourn} \& {Shanks}}{{Whitbourn} \&
  {Shanks}}{2014}]{WhitbournShanks14}
{Whitbourn} J.~R.,  {Shanks} T.,  2014, \mn@doi [\mnras]
  {10.1093/mnras/stt2024}, \href
  {http://adsabs.harvard.edu/abs/2014MNRAS.437.2146W} {437, 2146}

\bibitem[\protect\citeauthoryear{{Wojtak}, {Knebe}, {Watson}, {Iliev},
  {He{\ss}}, {Rapetti}, {Yepes}  \& {Gottl{\"o}ber}}{{Wojtak}
  et~al.}{2014}]{Wojtak14}
{Wojtak} R.,  {Knebe} A.,  {Watson} W.~A.,  {Iliev} I.~T.,  {He{\ss}} S.,
  {Rapetti} D.,  {Yepes} G.,   {Gottl{\"o}ber} S.,  2014, \mn@doi [\mnras]
  {10.1093/mnras/stt2321}, \href
  {http://adsabs.harvard.edu/abs/2014MNRAS.438.1805W} {438, 1805}

\bibitem[\protect\citeauthoryear{{Wu}, {Hahn}, {Evrard}, {Wechsler}  \&
  {Dolag}}{{Wu} et~al.}{2013}]{Wu13bv}
{Wu} H.-Y.,  {Hahn} O.,  {Evrard} A.~E.,  {Wechsler} R.~H.,   {Dolag} K.,
  2013, \mn@doi [\mnras] {10.1093/mnras/stt1582}, \href
  {http://adsabs.harvard.edu/abs/2013MNRAS.436..460W} {436, 460}

\bibitem[\protect\citeauthoryear{{Zehavi}, {Riess}, {Kirshner}  \&
  {Dekel}}{{Zehavi} et~al.}{1998}]{Zehavi98}
{Zehavi} I.,  {Riess} A.~G.,  {Kirshner} R.~P.,   {Dekel} A.,  1998, \mn@doi
  [\apj] {10.1086/306015}, \href
  {http://adsabs.harvard.edu/abs/1998ApJ...503..483Z} {503, 483}

\bibitem[\protect\citeauthoryear{{Zhang}, {Childress}, {Davis}, {Karpenka},
  {Lidman}, {Schmidt}  \& {Smith}}{{Zhang} et~al.}{2017}]{Zhang17}
{Zhang} B.~R.,  {Childress} M.~J.,  {Davis} T.~M.,  {Karpenka} N.~V.,  {Lidman}
  C.,  {Schmidt} B.~P.,   {Smith} M.,  2017, preprint, \href
  {http://adsabs.harvard.edu/abs/2017arXiv170607573Z} {} (\mn@eprint {arXiv}
  {1706.07573})

\bibitem[\protect\citeauthoryear{{de Vaucouleurs} \& {Bollinger}}{{de
  Vaucouleurs} \& {Bollinger}}{1979}]{deVaucouleurs79}
{de Vaucouleurs} G.,  {Bollinger} G.,  1979, \mn@doi [\apj] {10.1086/157405},
  \href {http://adsabs.harvard.edu/abs/1979ApJ...233..433D} {233, 433}

\makeatother
\end{thebibliography}

\bsp	
\label{lastpage}
\end{document}